\documentclass[aps,prd,onecolumn,superscriptaddress,showpacs]{revtex4}

\def\be{\begin{equation}}
\def\ee{\end{equation}}
\def\ba{\begin{eqnarray}}
\def\ea{\end{eqnarray}}

\begin{document}

\title{Can one detect new physics in $I=0$ and/or $I=2$ contributions
to the decays $B \rightarrow \pi \pi$?} 

\author{Seungwon Baek}
\affiliation{Physique des Particules,
        Universit\'{e} de Montr\'{e}al,
	C.P.\ 6128, succ.\ centre-ville,
	Montr\'{e}al, QC, Canada H3C 3J7}
\author{F.\ J.\ Botella}
\affiliation{Departament de F\'{\i}sica Te\`{o}rica and IFIC,
        Universitat de Val\`{e}ncia-CSIC,
        E-46100, Burjassot, Spain}
\author{David London}
\affiliation{Physique des Particules,
        Universit\'{e} de Montr\'{e}al,
	C.P.\ 6128, succ.\ centre-ville,
	Montr\'{e}al, QC, Canada H3C 3J7}
\author{Jo\~{a}o P.\ Silva}
\affiliation{Centro de F\'{\i}sica Te\'{o}rica de Part\'{\i}culas,
	Instituto Superior T\'{e}cnico,
	P-1049-001 Lisboa, Portugal}
\affiliation{Instituto Superior de Engenharia de Lisboa,
	Rua Conselheiro Em\'{\i}dio Navarro,
	1900 Lisboa, Portugal}

\date{\today}

\begin{abstract}
We study the effects of new-physics contributions to $B \rightarrow
\pi \pi$ decays, which can be parametrized as four new complex
quantities. A simple analysis is provided by utilizing the
reparametrization invariance of the decay amplitudes. We find that six
quantities can be reabsorbed into the definitions of Standard
Model-like parameters. As a result, the usual isospin analysis
provides only two constraints on new physics which are independent of
estimates for the Standard Model contributions. In particular, we show
that one is not sensitive to new physics affecting the $I=0$
amplitudes. On the other hand, $I=2$ new physics can be detected, and
its parameters can be measured by using independent determinations of
the weak phases. We obtain constraints on these new-physics parameters
through a fit to the current experimental data.
\end{abstract}


\maketitle

\section{\label{sec:intro}Introduction}

The purpose of $B$-physics experiments is the detection of new
physics.  Because CP violation appears in the Standard Model (SM)
through one single irremovable phase in the Cabibbo-Kobayashi-Maskawa
(CKM) matrix \cite{CKM}, early strategies involved determining the
various incarnations of this phase ($\beta$, $\gamma$, or $\alpha
\equiv \pi - \beta - \gamma$), looking for discrepancies.  Several
techniques were proposed to sidestep the need to deal with the
amplitude magnitudes and with the CP-even strong phases, since these
are affected by uncertain hadronic matrix elements -- reviews can be
found, for example, in \cite{BLS,CKMfitter-04,Prague04}.

In one such proposal, due to Gronau and London, one uses the isospin
symmetry between different $B \rightarrow \pi \pi$ decays \cite{GL}.
Their proposal can be worded in several different ways.  We may take
it as a measurement of $\beta + \gamma$, to be compared with the
values allowed for this quantity by current CKM constraints on the
Wolfenstein $\rho$--$\eta$ plane \cite{Wolf83}; we may use the
measurement of $\beta$ from $B_d \rightarrow \psi K$ decays, and view
this as a measurement of $\gamma$; or, one may take $\gamma_{\rm ckm}$
and $\beta_{\rm ckm}$ from the fit to the $\rho$--$\eta$ plane,
looking for inconsistencies in the overall fit of the SM parameters
(including all CP-odd and CP-even quantities) to the experimental
observables in $B \rightarrow \pi \pi$ decays.

In this article, we follow the last approach with respect to the weak
phases (dropping the subscript ``ckm''), but we will consider the most
general type of new physics that could affect these decays.  Our
objective is to find which types of new physics can be probed in $B
\rightarrow \pi \pi$ decays without making any assumptions about the
hadronic matrix elements of the SM contributions to these decays, and
which cannot.  We show that:
\begin{enumerate}
\item there are only two probes of new physics in $I=2$ contributions:
one probes the presence of a new weak phase in $A_2$; the other
compares the value of $\gamma_{\pi \pi}$ extracted from the isospin
analysis with that obtained independently through CKM unitarity or
some other decay;
\item one \textit{cannot} probe for new physics in $I=0$
contributions.
\end{enumerate}
We show how these conclusions follow simply from the
``reparametrization invariance'' introduced by two of us (Botella and
Silva) in \cite{reparametrization}. In addition, if a new weak phase
in $A_2$ is seen, we show that it is possible to {\it measure} the
new-physics parameters using independent determinations of the weak
phases.

In section~\ref{sec:repara}, we explain the generic features of
``reparametrization invariance'' relevant for this problem.  In
section~\ref{sec:parametrize}, we perform a general analysis of the $B
\rightarrow \pi \pi$ decays valid in the presence of new physics and
we prove that the conclusions announced above follow simply from
reparametrization invariance.  In section~\ref{sec:BL} we perform a
fit of the relevant new-physics parameters to the current experimental
data. These constraints on new physics do \textit{not} depend on any
assumptions about the SM contributions, which are also independently
extracted from our fit.  We present our conclusions in
section~\ref{sec:conclusions}.

\section{\label{sec:repara}Consequences of reparametrization invariance}

Let us consider the decay of a $B$ meson into some specific final
state $f$.  For the moment, $B$ stands for $B^+$, $B_d^0$ or $B_s^0$.
When discussing generic features of the decay amplitudes without
reference to any particular model, it has become commonplace to
parametrize the decay amplitudes as
\ba
A_f
&=&
M_1 e^{i \phi_{A1}} e^{i \delta_1} + M_2 e^{i \phi_{A2}} e^{i \delta_2},
\label{A_f}
\\
\bar A_{\bar f}
&=&
M_1 e^{- i \phi_{A1}} e^{i \delta_1} + M_2 e^{- i \phi_{A2}} e^{i \delta_2},
\label{Abar_fbar}
\ea
where $\phi_{A1}$ and $\phi_{A2}$ are two CP-odd weak phases; $M_1$
and $M_2$ are the magnitudes of the corresponding terms; and
$\delta_1$ and $\delta_2$ are the corresponding CP-even strong phases
\cite{detail}.  These expressions apply to the decays of a (neutral or
charged) $B$ meson into the final state $f$ and the charge-conjugated
decay, respectively.  For the decay of a neutral $B$ meson into a CP
eigenstate with CP eigenvalue $\eta_f = \pm 1$, the right-hand-side of
Eq.~(\ref{Abar_fbar}) appears multiplied by $\eta_f$.

As shown in reference~\cite{reparametrization}, the fact that any
third weak phase may be written in terms of the first two means that
one may write any amplitude, with an arbitrary number $N$ of distinct
weak phases, in terms of only two.  Indeed,
\be
A_f = 
\tilde{M}_1 e^{i \phi_{A1}} e^{i \tilde{\delta}_1}
+
\tilde{M}_2 e^{i \phi_{A2}} e^{i \tilde{\delta}_2}
+
\sum_{k=3}^{N} 
\tilde{M}_k e^{i \phi_{Ak}} e^{i \tilde{\delta}_k}
\label{master-A}
\ee
\textit{and}
\be
\bar A_{\bar f} = 
\tilde{M}_1 e^{-i \phi_{A1}} e^{i \tilde{\delta}_1}
+
\tilde{M}_2 e^{-i \phi_{A2}} e^{i \tilde{\delta}_2}
+
\sum_{k=3}^{N} 
\tilde{M}_k e^{-i \phi_{Ak}} e^{i \tilde{\delta}_k}
\label{master-Abar}
\ee
may be written as in Eqs.~(\ref{A_f}) and (\ref{Abar_fbar}),
respectively, through the choices
\ba
M_1 e^{i \delta_1} &=&
\tilde{M}_1 e^{i \tilde{\delta}_1}
+
\sum_{k=3}^{N}
a_k 
\tilde{M}_k  e^{i \tilde{\delta}_k},
\nonumber\\
M_2 e^{i \delta_2} &=&
\tilde{M}_2 e^{i \tilde{\delta}_2}
+
\sum_{k=3}^{N}
b_k 
\tilde{M}_k  e^{i \tilde{\delta}_k},
\ea
with
\begin{eqnarray}
a_k &=& \frac{\sin{(\phi_{Ak} - \phi_{A2})}}{\sin{(\phi_{A1} - \phi_{A2})}},
\nonumber\\
b_k &=& \frac{\sin{(\phi_{Ak} - \phi_{A1})}}{\sin{(\phi_{A2} - \phi_{A1})}}.
\label{ak_bk}
\end{eqnarray}
Notice that, in addition, the phases $\phi_{A1}$ and $\phi_{A2}$ may
be chosen completely at will.  This property, which we refer to as
``reparametrization invariance'', has very unusual consequences, which
were explored at length in \cite{reparametrization}.

Sometimes it is useful to consider the sums of all new contributions
to $B$ and $\overline{B}$ decays,
\begin{eqnarray}
N &=&
\sum_{k=3}^{N} 
\tilde{M}_k e^{i \phi_{Ak}} e^{i \tilde{\delta}_k},
\nonumber\\
\bar N &=&
\sum_{k=3}^{N} 
\tilde{M}_k e^{- i \phi_{Ak}} e^{i \tilde{\delta}_k}.
\end{eqnarray}
With this notation, the proof that we may use only two weak phases as
our basis follows simply from
\begin{eqnarray}
N & = &
N_{\phi_{A1}} e^{i \phi_{A1}} + N_{\phi_{A2}} e^{i \phi_{A2}},
\label{redefine-N}
\\
\bar N & = &
N_{\phi_{A1}} e^{-i \phi_{A1}} + N_{\phi_{A2}} e^{-i \phi_{A2}},
\label{redefine-Nbar}
\end{eqnarray}
where
\ba
N_{\phi_{A1}} &=&
\frac{N e^{-i \phi_{A2}} - \bar N e^{i \phi_{A2}}
}{2 i \sin{(\phi_{A1} - \phi_{A2})}}
\equiv
\sum_{k=3}^{N}
a_k 
\tilde{M}_k  e^{i \tilde{\delta}_k},
\nonumber\\
N_{\phi_{A2}} &=&
\frac{N e^{-i \phi_{A1}} - \bar N e^{i \phi_{A1}}
}{2 i \sin{(\phi_{A2} - \phi_{A1})}}
\equiv
\sum_{k=3}^{N}
b_k 
\tilde{M}_k  e^{i \tilde{\delta}_k}.
\label{N1N2}
\ea
Notice that, as required, the \textit{same} complex numbers
$N_{\phi_{A1}}$ and $N_{\phi_{A2}}$ appear in Eqs.~(\ref{redefine-N})
and (\ref{redefine-Nbar}).  Said otherwise, $N_{\phi_{A1}}$ and
$N_{\phi_{A2}}$ carry only magnitudes and CP-even phases, since the
CP-odd phases, $\phi_{A1}$ and $\phi_{A2}$, have been factored out
explicitly in Eqs.~(\ref{redefine-N}) and (\ref{redefine-Nbar}).

\section{\label{sec:parametrize}Parametrizing the
$B \rightarrow \pi \pi$ decay amplitudes}

We may parametrize the $B \rightarrow \pi \pi$ decay amplitudes
according to the isospin of the final state as
\ba
- \sqrt{2} A(B^+ \rightarrow \pi^+ \pi^0)
=
- \sqrt{2}
A_{+0} &=& 3 A_2,
\nonumber\\
-  A(B^0 \rightarrow \pi^+ \pi^-) = 
- A_{+-} &=& A_2 + A_0,
\nonumber\\
- \sqrt{2} A(B^0 \rightarrow \pi^0 \pi^0)
=
- \sqrt{2}
A_{00} &=& 2 A_2 - A_0,
\label{physical-A}
\ea
and
\ba
- \sqrt{2} A(B^- \rightarrow \pi^- \pi^0)
=
- \sqrt{2}
\bar A_{+0} &=& 3 \bar A_2,
\nonumber\\
-  A(\overline{B^0} \rightarrow \pi^+ \pi^-) = 
- 
\bar A_{+-} &=& \bar A_2 + \bar A_0,
\nonumber\\
- \sqrt{2} A(\overline{B^0} \rightarrow \pi^0 \pi^0)
=
- \sqrt{2}
\bar A_{00} &=& 2 \bar A_2 - \bar A_0.
\label{physical-Abar}
\ea
In writing Eqs.~(\ref{physical-A}) and (\ref{physical-Abar}), some
coefficients and signs have been absorbed into the definition of the
amplitudes for $I=0$ ($A_0$ and $\bar A_0$) and $I=2$ ($A_2$ and $\bar
A_2$); this choice is not universal and great care should be exercised
when comparing with other sources.

The right-hand-sides of Eqs.~(\ref{physical-A}) and
(\ref{physical-Abar}) contain seven independent parameters: four
magnitudes ($|A_2|$, $|\bar A_2|$, $|A_0|$, and $|\bar A_0|$); and
three relative phases ($\bar \delta_2 - \delta_2$, $\bar \delta_0 -
\delta_0$, and $\delta_2 - \delta_0$).  An overall phase can be
rotated away.  These seven quantities may be extracted from
experiments detecting the average branching ratios ($B_{+0}$,
$B_{+-}$, and $B_{00}$), the direct CP violation ($C_{+0}$, $C_{+-}$,
and $C_{00}$), and the interference CP violation ($S_{+-}$ and
$S_{00}$) of $B \rightarrow \pi \pi$ decays, where the sub-indices
refer to the charges of the physical pions in the final state.  It
turns out that $S_{00}$ may be written as a function of the other
observables, up to discrete ambiguities.  Therefore, there are seven
independent measurements in $B \rightarrow \pi \pi$ decays, allowing
the determination of the seven physical parameters present on the
right-hand-sides of Eqs.~(\ref{physical-A}) and (\ref{physical-Abar}).

A different decomposition is sometimes utilized within the SM.  This
is related to a diagrammatic analysis and it involves two weak phases
($\beta$ and $\gamma$) which appear naturally within the SM:
\ba
- \sqrt{2} A_{+0} &=& \left( T + C \right) e^{i \gamma},
\nonumber\\
- A_{+-} &=& T e^{i \gamma} + P  e^{- i \beta},
\nonumber\\
- \sqrt{2} A_{00} &=& C e^{i \gamma} - P  e^{- i \beta}.
\label{Apipi-SM-TPC}
\ea
Here $T$, $C$, and $P$ contain only magnitudes and CP-even (strong)
phases.  Similar relations hold for the conjugated (barred)
amplitudes, by changing the signs of the CP-odd phases $\gamma$ and $-
\beta$.  The relation between the two decompositions is
\ba
A_2 &=& \frac{1}{3} (T + C) e^{i \gamma},
\nonumber\\
\bar A_2 &=& \frac{1}{3} (T + C) e^{-i \gamma},
\nonumber\\
A_0 &=& \frac{1}{3} (2 T - C) e^{i \gamma} + P e^{-i \beta},
\nonumber\\
\bar A_0 &=& \frac{1}{3} (2 T - C) e^{-i \gamma} + P e^{i \beta}.
\label{SM-TPC}
\ea
For simplicity,
in writing Eqs.~(\ref{SM-TPC}) we have neglected the SM
electroweak penguin contributions, 
but these can be included in a straightforward way by shifting gamma
roughly by $1.5^\circ$,
following references \cite{PEW}.

The impact of a generic new-physics model in $B \rightarrow \pi \pi$
decays will show up in both $I=0$ and $I=2$ amplitudes, with a variety
of weak phases. This can be parametrized as
\ba
A_2 &=& \frac{1}{3} (T + C) e^{i \gamma} + N_2,
\nonumber\\
\bar A_2 &=& \frac{1}{3} (T + C) e^{-i \gamma} + \bar N_2,
\nonumber\\
A_0 &=& \frac{1}{3} (2 T - C) e^{i \gamma} + P e^{-i \beta} + N_0,
\nonumber\\
\bar A_0 &=& \frac{1}{3} (2 T - C) e^{-i \gamma} + P e^{i \beta} + \bar N_0,
\label{A0A2-with-new}
\ea
where $N_0$, $\bar N_0$, $N_2$, and $\bar N_2$ are complex numbers.
We may use the consequences of reparametrization invariance in
Eqs.~(\ref{redefine-N})--(\ref{N1N2}) in order to rewrite
Eqs.~(\ref{A0A2-with-new}) as
\ba
A_2 &=& \frac{1}{3} (t + c) e^{i \gamma} + N_{2,o}\ ,
\nonumber\\
\bar A_2 &=& \frac{1}{3} (t + c) e^{-i \gamma} + N_{2,o}\ ,
\nonumber\\
A_0 &=& \frac{1}{3} (2 t - c) e^{i \gamma} + p e^{-i \beta},
\nonumber\\
\bar A_0 &=& \frac{1}{3} (2 t - c) e^{-i \gamma} + p e^{i \beta}.
\label{small-tpc}
\ea
Here
\ba
t + c &=& T + C + 3 N_{2,\gamma}\ ,
\label{relate-t+c}\\
2 t - c &=& 2 T- C + 3 N_{0,\gamma}\ ,
\label{relate-2t-c}\\
p &=& P + N_{0,-\beta}\ ,
\label{relate-p}
\ea
where
\ba
N_{2, \gamma} &=&
i \frac{\bar N_2 - N_2}{2 \sin{\gamma}},
\nonumber\\
N_{2, o} &=&
\frac{\bar N_2 + N_2}{2}
- i \frac{\bar N_2 - N_2}{2 \tan{\gamma}},
\nonumber\\
N_{0,\gamma} &=&
\frac{\bar N_0 + N_0}{2} \frac{\sin{\beta}}{\sin{(\beta + \gamma)}}
+ i \frac{\bar N_0 - N_0}{2} \frac{\cos{\beta}}{\sin{(\beta + \gamma)}},
\nonumber\\
N_{0,-\beta} &=&
\frac{\bar N_0 + N_0}{2} \frac{\sin{\gamma}}{\sin{(\beta + \gamma)}}
- i \frac{\bar N_0 - N_0}{2} \frac{\cos{\gamma}}{\sin{(\beta + \gamma)}},
\label{N-phases}
\ea
are obtained from Eqs.~(\ref{redefine-N})--(\ref{N1N2}) with $\{
\phi_{A1}, \phi_{A2} \} = \{ \gamma, 0 \}$ for the $I=2$
contributions, and with $\{ \phi_{A1}, \phi_{A2} \} = \{ \gamma,
-\beta \}$ for the $I=0$ contributions.

We stress that our choice of $\{ \phi_{A1}, \phi_{A2} \} = \{ \gamma,
0 \}$ for the $I=2$ contributions is not mandatory.  We could equally
well have chosen a more general basis $\{ \phi_{A1}, \phi_{A2} \} = \{
\gamma, \phi \}$, as long as the phase $\phi$ was \textit{known} and
did not have to be fitted for \cite{general}.  For example, we could
take $\phi = 5^\circ$, or $\phi = 10^\circ$, or even $\phi = \beta$,
with $\beta$ determined from $B_d \rightarrow \psi K$ decays.

The main results of our paper arise by comparing
Eqs.~(\ref{small-tpc}), valid in the presence of generic new-physics
contributions to $B \rightarrow \pi \pi$ decays, with
Eqs.~(\ref{SM-TPC}), valid within the SM.  First, we notice that the
expressions for $A_0$ and $\bar A_0$ have exactly the same form in
Eqs.~(\ref{SM-TPC}) and in Eqs.~(\ref{small-tpc}).  This means that,
without specific assumptions made about the hadronic matrix elements
involved in the SM contributions $T$, $C$, and $P$, the measurements
of $A_0$ and $\bar A_0$ cannot be used to test for the presence of new
physics in $I=0$ (or lack thereof).  This is one of our main points.
It is \textit{impossible} to detect new physics in $I=0$ without
specific assumptions about the hadronic matrix elements involved in
the SM contributions. Note that the impossibility of detecting $I=0$
new physics has long been suspected; reparametrization invariance
offers a {\it proof} of this fact.

Conversely, if one makes assumptions about the quantities involved in
the SM contributions $2T-C$ and/or $P$, then the deviations
$(2t-c)_{\rm exp} - (2T-C)$ and $p_{\rm exp} - P$ \textit{can} indeed
be used to probe the $I=0$ contributions $N_{0, \gamma}$ and $N_{0,
-\beta}$, respectively.  This contradicts an analysis performed
earlier by two of us (Baek and London) in references~\cite{BL1,BL2}.
The imprecision had to do with a very subtle question related to
rephasing.  It is only in the language of reparametrization invariance
that this issue becomes simple to understand, illustrating how
powerful reparametrization invariance is as a tool to organize the
new-physics contributions.

Second, we notice that the expressions for $A_2$ and $\bar A_2$ do not
have the same form in Eqs.~(\ref{SM-TPC}) and in
Eqs.~(\ref{small-tpc}).  One piece of the new-physics contribution,
$N_{2, \gamma}$, can indeed be reabsorbed into the definition of
$t+c$, as in Eq.~(\ref{relate-t+c}).  (As with the $I=0$
contributions, the presence of the new $I=2$ contribution $N_{2,
\gamma}$ may only be tested for under specific assumptions for the SM
contributions to $T+C$.)  But the other piece, $N_{2, o}$, cannot be
reabsorbed by a redefinition of SM-like parameters.  This means that
the presence of some types of new physics in $I=2$ \textit{can} be
detected, even without specific assumptions made about the hadronic
matrix elements involved in the SM contributions $T$ and $C$.  Because
$N_{2, o}$ is a complex number, we expect two such tests; these are
related with the magnitude of $N_{2, o}$, and (once this magnitude is
nonzero) with the difference between its (strong) phase and that of
$t+c$.

To understand the first test, let us start by considering the case in
which the (strong) phase of $N_{2, o}$ coincides with that of $t+c$,
$\delta_{t+c}$.  In that case the $I=2$ amplitudes may be written as
\ba
A_2 &=& e^{i \delta_{t+c}}
\left[ \frac{1}{3} |t+c| e^{i \gamma} + |N_{2,o}|
\right]
= e^{i \delta_{t+c}} e^{i \gamma_{\pi \pi}} |A_2|,
\nonumber\\
\bar A_2 &=& e^{i \delta_{t+c}}
\left[ \frac{1}{3} |t+c| e^{- i \gamma} + |N_{2,o}|
\right]
= e^{i \delta_{t+c}} e^{- i \gamma_{\pi \pi}} |A_2|,
\label{only-tilde-gamma}
\ea
where
\be
\tan{\gamma_{\pi \pi}} =
\frac{\sin{\gamma}}{\cos{\gamma} + 3 \frac{|N_{2,o}|}{|t+c|}}.
\ee
This type of new physics will be seen as a difference between the
phase $\gamma_{\pi \pi}$ obtained from the isospin analysis of $B
\rightarrow \pi \pi$ decays and the phase $\gamma_{\rm ckm}$ obtained
from the current CKM constraints on the Wolfenstein $\rho$--$\eta$
plane.  Naturally, this signal of new physics disappears as $N_{2,o}$
vanishes.  Moreover, in this case, because the same $|A_2|$ appears on
both lines of Eq.~(\ref{only-tilde-gamma}), $|\bar A_{+0}|^2 -
|A_{+0}|^2 \propto |\bar A_2|^2 - |A_2|^2 =0$, and there is no direct
CP violation in $B^\pm \rightarrow \pi^\pm \pi^0$ decays.  So, the
(one) test of new physics possible when $C_{+0} = 0$ is
\be
\left|
\frac{N_{2,o}}{t+c}
\right|
=
\frac{\sin(\gamma_{\rm ckm} - \gamma_{\pi \pi})}{3
\sin{\gamma_{\pi \pi}}}.
\ee

The second test on $N_{2,o}$ arises if it carries a strong phase which
differs from $\delta_{t+c}$.  In that case $|\bar A_2|$ differs from
$|A_2|$, and this will be reflected in the appearance of direct CP
violation in $B^\pm \rightarrow \pi^\pm \pi^0$ decays.

In both cases, if we take the values of $\gamma$ and $-\beta$ from
independent measurements, the number of observables in $B \rightarrow
\pi \pi$ decays is equal to the number of theoretical parameters.
Thus, it is not only possible to detect a nonzero $N_{2,o}$; one can
also measure its parameters. Up to now, this has not been realized; as
above, it is only by using reparametrization invariance that one sees
this.

We conclude that there are only two independent tests for new physics
in $B \rightarrow \pi \pi$ decays which do not depend on hadronic
estimates for the SM contributions.  New physics in $I=0$
contributions and $N_{2, \gamma}$ pieces in $I=2$ cannot be tested
for.  In contrast, $N_{2, o}$ contributions can be tested for, and
they appear as $\gamma_{\pi \pi} - \gamma_{\rm ckm} \neq 0$, or
$C_{+0} \neq 0$. In addition, if the weak phases are assumed to be
known independently, one can measure the parameters of $N_{2, o}$.
Further tests and measurements are possible if one makes specific
assumptions about the hadronic matrix elements of the SM.

\section{\label{sec:BL}Constraining new-physics contributions with current
data}

The present $B \rightarrow \pi \pi$ measurements are detailed
in Table~\ref{T:data}.
%
\begin{table}[h]
\caption{\label{T:data}Branching ratios, direct CP asymmetries $C_f$,
and interference CP asymmetries $S_f$ (if applicable) for the three $B
\rightarrow \pi\pi$ decay modes. Data comes from
Refs.~\cite{BRs,ACPs,pi+pi-}; averages (shown) are taken from
Ref.~\cite{HFAG}.}
\begin{ruledtabular}
\begin{tabular}{lccc}
& $BR[10^{-6}]$ & $C_f$  & $S_f$ \\
\hline
$B^+ \rightarrow \pi^+ \pi^0$ & $5.5 \pm 0.6$ & $0.02 \pm 0.07$ & \\
$B^0 \rightarrow \pi^+ \pi^-$ & $4.6 \pm 0.4$ & $-0.37 \pm 0.10$ & $-0.50 \pm
    0.12$ \\
$B^0 \rightarrow \pi^0 \pi^0$ & $1.51 \pm 0.28$ & $-0.28 \pm 0.39$ \\
\end{tabular}
\end{ruledtabular}
\end{table}
%
The phase $\beta$ is taken from the measurements of interference CP
violation in $B \rightarrow \psi K$ decays: $\sin{2 \beta} = 0.725 \pm
0.037$ \cite{beta}.  Thus, $2 \beta$ is determined up to a twofold
ambiguity.  We assume that $\beta \sim 23.5^\circ$, in agreement with
the SM.  The value of $\gamma$ is taken from independent measurements
\cite{CKMfitter}.  For the purposes of the fit, we assume symmetric
errors, and take $\gamma = (58.2 \pm 6.0)^\circ$.

Using the independent determinations of the SM CP phases, along with
the latest $B \rightarrow \pi \pi$ measurements, we obtain the values
for the isospin amplitudes.  The fit to present data yields four
solutions, presented in Table~\ref{T:fit_isospin}.
%
\begin{table}[h]
\caption{\label{T:fit_isospin}Results of a fit of the isospin
amplitudes to current $B \rightarrow \pi \pi$ data.  We have factored
out the (unphysical) overall phase $\bar \delta_0$.  The magnitudes
are measured in $eV$ and the phases in degrees.}
\begin{ruledtabular}
\begin{tabular}{ccccccc}
 $|A_2|$  & $|A_0|$  & $|\bar{A}_2|$  & $|\bar{A}_0|$  & 
 $\delta_2-\bar \delta_0$ & $\delta_0-\bar \delta_0$ & 
 $\bar \delta_2 -\bar \delta_0$ \\
\hline
   11.4 $\pm$ 0.7 & 6.8 $\pm$ 1.6 & 11.2 $\pm$ 0.7& 19.3 $\pm$ 2.0 &
   $-$35.1 $\pm$ 80.1& $-$35.1 $\pm$ 134 & $-$59.5 $\pm$ 9.3 \\
   11.4 $\pm$ 0.7 & 6.8 $\pm$ 1.6 & 11.2 $\pm$ 0.7& 19.3 $\pm$ 2.0 &
   7.6 $\pm$ 80.1& 7.6 $\pm$ 134 & 59.5 $\pm$ 9.3 \\
   11.4 $\pm$ 0.7 & 6.8 $\pm$ 1.6 & 11.2 $\pm$ 0.7& 19.3 $\pm$ 2.0 &
   79.8 $\pm$ 80.1& 79.8 $\pm$ 134 & $-$59.5 $\pm$ 9.3 \\
   11.4 $\pm$ 0.7 & 6.8 $\pm$ 1.6 & 11.2 $\pm$ 0.7& 19.3 $\pm$ 2.0 &
   122 $\pm$ 80.1& 122 $\pm$ 134 & 59.5 $\pm$ 9.3 \\
\end{tabular}
\end{ruledtabular}
\end{table}
%
We get $\chi^2_{min}/d.o.f. = 0.0049/0$, which is larger than
expected.  This occurs because the current data are slightly
inconsistent with the isospin $\{A_0, A_2\}$ description.  Indeed, we
have for the central values
\be
\cos(\delta_2 - \delta_0) = 
\frac{{2 \over 3} |A_{+0}|^2 + |A_{+-}|^2 -2 |A_{00}|^2}{
2 \sqrt{2} |A_{+0}| |A_0|}
= 1.07,
\ee
where $|A_0|$ is given by
\be
|A_0|^2 = \frac{2}{3}
\left( 
-\frac{2}{3} |A_{+0}|^2 + |A_{+-}|^2 + |A_{00}|^2
\right).
\ee
This explains why our fit gives the same values for $\delta_2$ and
$\delta_0$.

We now wish to perform the fit in the notation of diagrammatic
amplitudes.  Using the rephasing freedom to set $\arg N_{2,0} = 0$, we
obtain the results in Table~\ref{T:fit_diagrammatic}.
%
\begin{table}[h]
\caption{\label{T:fit_diagrammatic}Results of a fit of the
diagrammatic amplitudes to current $B \rightarrow \pi \pi$ data.  We
have factored out the (unphysical) overall phase $\delta_{N_{2,0}} =
\arg N_{2,0}$.  The magnitudes are measured in $eV$ and the phases in
degrees.}
\begin{ruledtabular}
\begin{tabular}{ccccccc}
$|t|$  & $|c|$  &  $|p|$ & $|N_{2,0}|$ & 
              $\delta_t - \delta_{N_{2,0}}$ &
              $\delta_c - \delta_{N_{2,0}}$ &
              $\delta_p - \delta_{N_{2,0}}$ \\
\hline
     6.1 $\pm$ 2.7 & 9.9 $\pm$ 13.7 & 12.9 $\pm$ 3.2 & 9.6 $\pm$ 6.5 &
     81.5 $\pm$ 70.5 & $-$40.5 $\pm$ 90.0& 22.3 $\pm$ 74.1 \\
      2.8 $\pm$ 2.6 & 19.8 $\pm$ 23.8 & 11.4 $\pm$ 6.6 & 13.2 $\pm$
      1.3 & 41 $\pm$ 108 & $-$174 $\pm$ 9 & $-$48.6 $\pm$ 64.2\\
     22.8 $\pm$ 4.0 & 18.2 $\pm$ 6.7 & 7.3 $\pm$ 6.5 & 2.7 $\pm$ 9.3 &
     $-$156 $\pm$ 52 & 155 $\pm$ 32 & 157 $\pm$ 20 \\
     19.6 $\pm$ 3.9 & 6.1 $\pm$ 22.4 & 6.4 $\pm$ 1.7 & 6.0 $\pm$ 8.4 &
     $-$19.1 $\pm 43.9$ & 68.9 $\pm$ 174 & $-$127 $\pm$ 35 \\
\end{tabular}
\end{ruledtabular}
\end{table}
%
We get $\chi^2_{min} = 0.0049$.

The results in Table~\ref{T:fit_diagrammatic} are related to those in
Table~\ref{T:fit_isospin} through
\ba
p &=&
\frac{\bar A_0\, e^{i \gamma} - A_0\, e^{-i \gamma}}{
2 i \sin{(\beta + \gamma)}},
\nonumber\\*[4mm]
t &=&
- \frac{\bar A_2  - A_2}{2 i \sin{\gamma}}
- \frac{\bar A_0\, e^{-i \beta} - A_0\, e^{i \beta}}{
2 i \sin{(\beta + \gamma)}},
\nonumber\\*[4mm]
c &=&
- 2\; \frac{\bar A_2  - A_2}{2 i \sin{\gamma}}
+ \frac{\bar A_0\, e^{-i \beta} - A_0\, e^{i \beta}}{
2 i \sin{(\beta + \gamma)}},
\nonumber\\*[4mm]
N_{2,0} &=& 
\frac{\bar A_2\, e^{i \gamma} - A_2\, e^{-i \gamma}}{
2 i \sin{\gamma}}.
\label{eq:analytic}
\ea
One could be worried by the fact that we have used the rephasing
freedom in order to set $\bar \delta_0 = 0$ when obtaining
Table~\ref{T:fit_isospin}, while we have used the rephasing freedom in
order to set $\arg{N_{2,0}} = 0$ in obtaining
Table~\ref{T:fit_diagrammatic}.  Nevertheless, both Tables contain
only rephasing-invariant quantities which, therefore, can be related.
It is easy to see how the rephasing freedom drops out from
Eqs.~(\ref{eq:analytic}) when one relates rephasing-invariant
quantities in both parametrizations.

We have also performed the fit of the current experimental
data to the SM,
obtained by setting $N_{2,0} = 0$.
The results are listed in 
Table~\ref{T:fit_SM}.
%
\begin{table}[h]
\caption{\label{T:fit_SM}Results of a fit of the SM diagrammatic
amplitudes to current $B \rightarrow \pi \pi$ data.  We have factored
out the (unphysical) overall phase $\delta_p$.  The magnitudes are
measured in $eV$ and the phases in degrees.}
\begin{ruledtabular}
\begin{tabular}{ccccc}
$|t|$  & $|c|$  &  $|p|$ & 
              $\delta_t - \delta_p$ & $\delta_c - \delta_p$ \\
\hline
     21.9 $\pm$ 1.1 & 18.7 $\pm$ 1.7 & 5.4 $\pm$ 1.5 & 55.6 $\pm$ 14.7
        & $-$11.9 $\pm$ 16.9\\
\end{tabular}
\end{ruledtabular}
\end{table}
%
We find $\chi^2_{min}/d.o.f. = 0.296/2$, meaning that, if one waives
any predictions for the hadronic matrix elements, then the SM provides
an excellent fit to the current data.

Notice that Table~\ref{T:fit_SM} only has one solution, while
Table~\ref{T:fit_diagrammatic} had four.  The reason is the following:
in the SM $\bar A_2 = A_2 e^{-2 i \gamma}$, or, in term of rephasing
invariant quantities,
\be
|\bar A_2|\, e^{i(\bar \delta_2 - \bar \delta_0)}
=
|A_2|\, e^{i(\delta_2 - \bar \delta_0)}\, e^{- 2 i \gamma}.
\label{check}
\ee
We can see that the third solution in Table~\ref{T:fit_diagrammatic}
is the one which best satisfies Eq.~(\ref{check}), giving the smallest
$\chi^2$ of all.

\section{\label{sec:conclusions}Conclusions}

We have considered the most general new-physics contributions to the
$I=0$ and $I=2$ amplitudes in $B \rightarrow \pi \pi$ decays, which
involve 4 new complex parameters $N_0$, $\bar N_0$, $N_2$, and $\bar
N_2$.  We have shown that $N_0$ and $\bar N_0$ may be absorbed by a
redefinition of the SM contributions to $B \rightarrow \pi \pi$
decays, as can $N_{2, \gamma}$, \textit{c.f.\/}
Eqs.~(\ref{relate-t+c})--(\ref{relate-p}).  This means that
new-physics contributions of this type -- and in particular, all
new-physics contributions to $I=0$ -- \textit{cannot} be detected
unless specific ranges are taken for the SM contributions.  In
contrast, $N_{2,o}$ allows for two tests for the new physics, related
to $C_{+0}$ and $\gamma_{\pi \pi} - \gamma_{\rm ckm}$.  These are the
only two probes of new physics in $B \rightarrow \pi \pi$ decays which
do not involve estimates of the SM hadronic matrix
elements. Furthermore, if one takes values for the weak phases from
independent determinations, the $B \rightarrow \pi \pi$ observables
allow one to measure the $N_{2,o}$ parameters. We have shown that all
of these conclusions follow simply from the reparametrization
invariance introduced in \cite{reparametrization}, thus illustrating
the power of this concept in providing a clear organization of the
new-physics contributions.

\begin{acknowledgments}
We thank Y.\ Nir, G.\ Raz, and L.\ Wolfenstein for discussions.  J.\
P.\ S.\ is extremely grateful to Y.\ Nir and to the Department of
Particle Physics of the Weizmann Institute of Science for their
excellent hospitality, while portions of the this work were made.  The
work of S.\ B.\ and D.\ L.\ is supported by NSERC of Canada.  F.\ J.\
B.\ is partially supported by the spanish M.\ E.\ C.\ under
FPA2002-00612 and HP2003-0079 (``Accion Integrada
hispano-portuguesa'').  J.\ P.\ S.\ is supported in part by the
Portuguese \textit{Funda\c{c}\~{a}o para a Ci\^{e}ncia e a Tecnologia}
(FCT) under the contract CFTP-Plurianual (777), and through the
project POCTI/37449/FNU/2001, approved by the Portuguese FCT and
POCTI, and co-funded by FEDER.
\end{acknowledgments}


\begin{thebibliography}{99}
\bibitem{general}
To be specific,
we may parametrize
%
\ba
N_2 &=& N_{2,\gamma}\, e^{i \gamma} + N_{2,\phi}\, e^{i \phi},
\nonumber\\
\bar N_2 &=& N_{2,\gamma}\, e^{- i \gamma} + N_{2,\phi}\, e^{- i \phi},
\ea
%
in which case the first two lines of Eq.~(\ref{N-phases}) become
%
\ba
N_{2, \gamma} &=&
- \frac{\bar N_2 + N_2}{2}
\frac{\sin{\phi}}{\sin{(\gamma - \phi)}}
+
i \frac{\bar N_2 - N_2}{2}
\frac{\cos{\phi}}{\sin{(\gamma - \phi)}},
\nonumber\\
N_{2, \phi} &=&
\frac{\bar N_2 + N_2}{2}
\frac{\sin{\gamma}}{\sin{(\gamma - \phi)}}
-
i \frac{\bar N_2 - N_2}{2}
\frac{\cos{\gamma}}{\sin{(\gamma - \phi)}}.
\ea
%
Clearly
%
\be
N_{2,\phi} = \frac{\sin{\gamma}}{\sin{(\gamma - \phi)}} N_{2,0}.
\ee
\end{thebibliography}
\end{document}